\documentclass[10pt]{article}

\newcommand{\tbsp}{\rule{0pt}{18pt}}

\usepackage{supertabular}
\usepackage{amsmath}
\usepackage{bm}
\usepackage[numbers,sort&compress]{natbib}
\begin{document}
\renewcommand{\thetable}{\Alph{table}}
\title{Multipole (E1, M1, E2, M2) transition wavelengths and rates 
between states with $n\leq 6$ in heliumlike
carbon, nitrogen, oxygen, neon, silicon, and argon}

\author{I. M. Savukov, W. R. Johnson, and U. I. Safronova\\
Department of Physics, 225 Nieuwland Science Hall \\
University of Notre Dame, Notre Dame, IN 46556}

\maketitle

\begin{abstract}
Transition wavelengths and rates are given for E1, E2, M1, and M2
transitions between singlet and triplet $S$, $P$, $D$, and $F$
states in heliumlike ions of astrophysical interest: carbon,
nitrogen, oxygen, neon, silicon, and argon. All possible
transitions between states with $n\leq 6$ are considered. Wave
functions and energies are calculated using the 
relativistic configuration-interaction
(CI) method including both Coulomb and Breit
interactions. For transitions to the ground state, the present 
theoretical wavelengths agree to five digits with precise measurements.
\end{abstract}

\newpage
\tableofcontents
\listoftables

\newpage

\section*{INTRODUCTION}
\addcontentsline{toc}{section}{INTRODUCTION}

In this paper, we present tables of energies and lifetimes of
states of heliumlike ions with $n \leq 6$ and $L\leq 3$ for ions
with $Z$ from 6 to 18 obtained from relativistic
configuration-interaction (CI) calculations.
In addition, we present tables of
electric dipole (E1), electric-quadrupole (E2), magnetic-dipole (M1),
and magnetic-quadrupole (M2) wavelengths and transition rates.

In a recent paper \cite{pap1}, we gave wavelengths
and transition rates for a limited set of E1 transitions
in heliumlike ions.  The aim of the
present paper is to extend the calculations to
give a more complete treatment of E1 transitions and to include
forbidden E2, M1, and M2 transitions as well.
In the most recent National Institute of Standards and Technology
(NIST) data compilation by \citet{nist96}, recommended
transition rates for heliumlike C, N, and O are given for relatively
few (23) forbidden transitions. By contrast, recommended values are
given for hundreds of allowed E1 transitions.
Among the forbidden transitions,
the $1\, ^1\! S_0$ - $2\, ^3\! S_1$ M1 transitions are the best studied,
 both theoretically
\cite{M1-69,M1-71,M1-85,PJS95,M1-98} and experimentally
\cite{M1Xe-89,M1Br-90,M1Nb-94,M1C-94}.
Relatively few results are available for M2
\cite{PJS95,M2-89,M2Ag-93} and E2 \cite{E2-80} transitions.

We evaluate 2172 E1, E2, M1, and M2 matrix elements
between states with $n \leq 6$ for C~V; the
corresponding numbers are 2178, 2190, 2224, 2265, and 2285
for N~VI, O~VII, Ne~IX, Si~XIII, and Ar~XVII, respectively.
The number of transitions is different for different ions since
we ignored those transitions with energy differences
less than 100 cm$^{-1}$.

\section*{METHOD}
\addcontentsline{toc}{section} {METHOD}
\subsection*{Wave Functions and Energies}
\addcontentsline{toc}{section} {Wave Functions and Energies}
The method used in the present calculation is described in detail
in  \citet{PJS95}. Briefly, relativistic CI calculations \cite{chen93,cheng94}
are used to obtain wave functions and energies.
The wave function describing a state with angular momentum $J, M$
in a two-electron ion is written  as
\begin{equation}
\Psi_{JM} = \sum_{i \le j} c_{ij} \Phi_{ij}(JM) ,  \label{eq1}
\end{equation}
where the quantities $c_{ij}$ are expansion coefficients and where
the configuration state vectors $\Phi_{ij}(JM)$ are given,
in second quantization, by
\begin{equation}
\Phi_{ij}(JM) = \eta_{ij} \sum_{m_i m_j} \langle j_i m_i , j_j m_j
| JM\rangle \; a^{\dagger}_i a^{\dagger}_j |0\rangle , \label{eq2}
\end{equation}
 with
\begin{equation}
\eta_{ij} = \left\{
\begin{array}{ll}
1,  & i\ne j ,\\ 1/{\sqrt 2}, & i = j .
\end{array}
\right.
  \label{eq3}
\end{equation}
The quantities $c_{ij}$, $\Phi_{ij}(JM)$, and $\eta_{ij}$ are
independent of magnetic quantum numbers. To construct a state of
even or odd parity, one requires the sum $l_i + l_j$  to be either
even or odd, respectively. From the symmetry properties of the
Clebsch-Gordan coefficients, it can be shown that
\begin{equation}
\Phi_{ij}(JM) = (-1)^{j_i + j_j + J + 1} \Phi_{ji}(JM) .
\label{eq4}
\end{equation}
This relation, in turn,  implies that $\Phi_{ii}(JM)$ vanishes
unless $J$ is even. The wave-function normalization condition has
the form
\begin{equation}
\langle\Psi_{JM}|\Psi_{JM}\rangle = \sum_{i \le j} c_{ij}^2  = 1 .
\label{eq5}
\end{equation}

Substituting $\Psi_{JM}$ into the Schr\"{o}dinger equation $(H_0
+V) \Psi_{JM} = E \Psi_{JM}$, one obtains the following set of
linear equations for the expansion coefficients $c_{ij}$:
\begin{equation}
  (\epsilon_i + \epsilon_j) c_{ij} + \sum_{k \le l}
 \eta_{ij} V(ij;kl)\, \eta_{kl} \, c_{kl} = E c_{ij} \,
  . \label{eq6}
\end{equation}
The potential matrix in Eq.~(\ref{eq6}) is
\begin{eqnarray}
V(ij;kl) &=& \sum_L (-1)^{j_j + j_k + L + J}
  \left\lbrace
  \begin{array}{ccc}
  j_i&j_j&J\\
  j_l&j_k&L
 \end{array}
  \right\rbrace
  X_L(ijkl) \nonumber \\ & +& \sum_L (-1)^{j_j + j_k + L}
  \left\lbrace
  \begin{array}{ccc}
  j_i&j_j&J\\
  j_k&j_l&L
  \end{array}
  \right\rbrace
  X_L(ijlk) ,
\label{eq7}
\end{eqnarray}
where the quantities $X_L(ijkl)$ are given by
\begin{equation}
X_L(ijkl) = (-1)^L \langle\kappa_i\|C_L\|\kappa_k\rangle
         \langle\kappa_j\|C_L\|\kappa_l\rangle R_L(ijkl) .\label{eq8}
\end{equation}
The coefficients $\langle\kappa_i\|C_L\|\kappa_j\rangle$ in the
above equation are reduced matrix elements of normalized spherical
harmonics, and the quantities $R_L(ijkl)$ are relativistic Slater
integrals defined in Refs.~\cite{chen93,PJS94}. In the present
calculation, where both Coulomb and Breit interactions are
included in the Hamiltonian,
\begin{equation}
X_L(ijkl) \rightarrow X_L(ijkl) + M_L(ijkl) + N_L(ijkl) +
O_L(ijkl) \, , \label{eq9}
\end{equation}
where $M_L(ijkl)$, $N_L(ijkl)$, and $O_L(ijkl)$ are the magnetic
integrals defined in Ref.~\cite{JBS88}.
We solve the eigenvalue problem (\ref{eq6}) to high
accuracy using the methods described by \citet{chen93}.
For the high-$n$ states considered herein, 
we closely follow the numerical procedures described in our
previous work on E1 transitions \cite{pap1}.

\subsection*{Transition Amplitudes and Rates}
\addcontentsline{toc}{section} {Transition Amplitudes and Rates}
Using the CI wave functions discussed above  for
both the initial and final states and carrying out the sums over
magnetic substates, one obtains the following expression for the
reduced $K$-pole matrix element for transitions between two states
$vw(J)$ and $v'w'(J')$ \cite{PJS95}:
\begin{eqnarray}
\lefteqn{ Z_K [v_{1}w_{1}(J)-v_{2}w_{2}(J')] =\sqrt{[J][J']}
\sum_{vw}\sum_{v'w'} {\cal S}^{J}(v_{1}w_{1},vw) {\cal
S}^{J'}(v_{2}w_{2},v'w') } \hspace{1.8in}  \nonumber \\ &&\times
(-1)^{K+j_{w}+j_{v'}} \left\{
\begin{array}{ccc}
J & J' & K \\ j_{v'} & j_{w} & j_{v}
\end{array}
\right\} Z_{K}(v'w) \delta _{vw'} ,  \label{eq10}
\end{eqnarray}
where $[J] = 2J+1$. The quantity ${\cal S}^{J}(v_{1}w_{1},vw)$ is
a symmetry coefficient defined by
\begin{equation}
{\cal S}^{J}(v_{1}w_{1},vw)=\eta _{v_{1}w_{1}}\left[ \delta
_{v_{1}v}\delta _{w_{1}w}+(-1)^{j_{v}+j_{w}+J+1}\delta
_{v_{1}w}\delta _{w_{1}v}\right] .
\end{equation}

We consider here E1 and E2 matrix elements in length and velocity
forms as well as  M1 and M2  matrix elements.  The E1 one-electron
reduced matrix elements $Z_{1}(vw)$ are given by Eqs.~(11--12)
in \citet{pap1}. The retarded E2,
M1, and M2 matrix elements are given by:
\begin{description}
\item[{\bf E2 length}]
\begin{eqnarray}
\lefteqn{
Z_{2}(vw) =\left\langle \kappa _{v}\left\| C_{2}\right\| \kappa
_{w}\right\rangle \frac{15}{k^{2}}\int_{0}^{\infty }dr \Biggl\{
j_{2}(kr)\left[ G_{v}(r)G_{w}(r)+F_{v}(r)F_{w}(r)\right] \Biggr. }
 \hspace{0em} \nonumber \\
&&\Biggl. +j_{3}(kr)\left[ \frac{\kappa _{v}-\kappa
_{w}}{3}\left[ G_{v}(r)F_{w}(r)+F_{v}(r)G_{w}(r)\right] +\left[
G_{v}(r)F_{w}(r)-F_{v}(r)G_{w}(r)\right] \right] \Biggr\} ,\label{eq11}
\end{eqnarray}
\item[{\bf E2 velocity}]
\begin{eqnarray}
\lefteqn{Z_{2}(vw) =\left\langle \kappa _{v}\left\| C_{2}\right\| \kappa
_{w}\right\rangle \frac{15}{k^{2}}\int_{0}^{\infty }dr
\Biggl\{ 2\frac{%
j_{2}(kr)}{kr}\left[ G_{v}(r)F_{w}(r)-F_{v}(r)G_{w}(r)\right]
\Biggr. }\hspace{6.5em} \nonumber\\
&&\Biggl. -\frac{\kappa _{v}-\kappa _{w}}{3}\left[ -j_{3}(kr)+\frac{3}{kr}%
j_{2}(kr)\right] \left[ G_{v}(r)F_{w}(r)+F_{v}(r)G_{w}(r)\right]
\Biggr\} , \label{eq12}
\end{eqnarray}
\item[{\bf M1}]
\begin{equation}\label{eq14}
Z_{1}(vw)=\left\langle -\kappa _{v}\left\| C_{1}\right\| \kappa
_{w}\right\rangle \frac{6}{\alpha k}\int_{0}^{\infty
}dr\frac{(\kappa _{v}+\kappa _{w})}{2}j_{1}(kr)\left[
G_{v}(r)F_{w}(r)+F_{v}(r)G_{w}(r)\right] ,
\end{equation}
\item[{\bf M2}]
\begin{equation}\label{eq13}
Z_{2}(vw)=\left\langle -\kappa _{v}\left\| C_{2}\right\| \kappa
_{w}\right\rangle \frac{30}{\alpha (k)^{2}}\int_{0}^{\infty
}dr\frac{(\kappa _{v}+\kappa _{w})}{3}j_{2}(kr)\left[
G_{v}(r)F_{w}(r)+F_{v}(r)G_{w}(r)\right]  .
\end{equation}
\end{description}
In the above equations, $\kappa _{v}$ is the angular momentum quantum number $[\kappa
_{v}=\mp (j_{v}+\frac{1}{2})$ for $j_{v}=(l_{v}\pm \frac{1}{2})]$,
and $k=\omega \alpha $, where $\omega =\varepsilon
_{w}-\varepsilon _{v}$ is the photon
energy and $\alpha $ is the fine-structure constant. The quantity
$C_{1q}(\hat{r})$ is a normalized spherical harmonic and $j_l(kr)$ is a
spherical Bessel function of order $l$. The functions $G_{a}(r)$ and
$F_{a}(r)$ are large- and small-component radial Dirac wave
functions, respectively.

The E1, E2, M1, and M2 transition probabilities $A_{FI}$~(s$^{-1}$)
are obtained in terms of line strengths $S_{FI}$~(a.u.)
and wavelength $\lambda$(\AA) as
\begin{eqnarray}
A^{E1}_{FI}&=&\frac{2.02613\times 10^{18}}{[J_{I}]\lambda^3}\ S^{E1}_{FI}, \\
A^{E2}_{FI}&=&\frac{1.11995\times 10^{18}}{[J_{I}]\lambda^5}\ S^{E2}_{FI}, \\
A^{M1}_{FI}&=&\frac{2.69735\times 10^{13}}{[J_{I}]\lambda^3}\ S^{M1}_{FI}, \\
A^{M2}_{FI}&=&\frac{1.49097\times 10^{13}}{[J_{I}]\lambda^5}\ S^{M2}_{FI}.
\end{eqnarray}

\section*{COMPARISON AND DISCUSSION}
\addcontentsline{toc}{section}{COMPARISON AND DISCUSSION}
\begin{table}
\begin{center}

\caption{Comparison  of wavelengths $\lambda$ (\AA),
line strengths $S$ (a.u.), and transition rates $A$ (s$^{-1}$)
with NIST data \cite{nist96}.
Numbers in brackets represent powers of 10.\label{tab-nist}
 }
\begin{tabular}{lllllllll}\hline
\multicolumn{1}{c}{} &
\multicolumn{1}{c}{Upper} &
\multicolumn{1}{c}{Lower} &
\multicolumn{2}{c}{$\lambda$ (\AA)} &
\multicolumn{2}{c}{$S$(a.u.)}&
\multicolumn{2}{c}{$A$ (s$^{-1}$)} \\
\multicolumn{1}{c}{Ion} &
\multicolumn{1}{c}{level} &
\multicolumn{1}{c}{level} &
\multicolumn{1}{c}{CI} &
\multicolumn{1}{c}{NIST} &
\multicolumn{1}{c}{CI} &
\multicolumn{1}{c}{NIST} &
\multicolumn{1}{c}{CI} &
\multicolumn{1}{c}{NIST} \\
\hline \multicolumn{9}{c}{M1 transitions} \\
C V  &$2\ ^3S_1$&$ 1\ ^1S_0$& 41.4717& 41.4721& 3.866[-7]& 3.853[-7] & 4.874[\hspace{0.35em}1]&  4.857[\hspace{0.35em}1]\\
N VI &$2\ ^3S_1$&$ 1\ ^1S_0$& 29.5341& 29.5343& 7.285[-7]& 7.30[-7]  & 2.543[\hspace{0.35em}2]&  2.55[\hspace{0.35em}2] \\
N VI &$3\ ^3S_1$&$ 1\ ^1S_0$& 25.0504& 25.0510& 1.830[-7]& 1.46[-7]  & 1.047[\hspace{0.35em}2]&  8.36[\hspace{0.35em}1] \\
N VI &$4\ ^3S_1$&$ 1\ ^1S_0$& 23.8270& 23.8277& 7.297[-8]& 6.91[-8]  & 4.850[\hspace{0.35em}1]&  4.59[\hspace{0.35em}1] \\
O VII&$2\ ^3S_1$&$ 1\ ^1S_0$& 22.0969& 22.1012& 1.257[-6]& 1.26[-6]  & 1.047[\hspace{0.35em}3]&  1.04[\hspace{0.35em}3] \\
O VII&$3\ ^3S_1$&$ 1\ ^1S_0$& 18.7269& 18.7307& 3.186[-7]& 2.61[-7]  & 4.362[\hspace{0.35em}2]&  3.57[\hspace{0.35em}2] \\
O VII&$4\ ^3S_1$&$ 1\ ^1S_0$& 17.8045& 17.8052& 1.274[-7]& 1.16[-7]  & 2.030[\hspace{0.35em}2]&  1.84[\hspace{0.35em}2] \\
\multicolumn{9}{c}{M2 transitions} \\
C V  &$2\ ^3P_2$&$ 1\ ^1S_0$& 40.7274& 40.7285& 9.860[-1]& 9.85[-1]  & 2.624[\hspace{0.35em}4]&  2.62[\hspace{0.35em}4] \\
C V  &$3\ ^3P_2$&$ 1\ ^1S_0$& 35.0693& 35.0703& 1.685[-1]& 1.65[-1]  & 9.475[\hspace{0.35em}3]&  9.30[\hspace{0.35em}3] \\
C V  &$4\ ^3P_2$&$ 1\ ^1S_0$& 33.4623& 33.4632& 5.962[-2]& 5.85[-2]  & 4.238[\hspace{0.35em}3]&  4.16[\hspace{0.35em}3] \\
C V  &$5\ ^3P_2$&$ 1\ ^1S_0$& 32.7716& 32.7726& 2.819[-2]& 2.73[-2]  & 2.224[\hspace{0.35em}3]&  2.15[\hspace{0.35em}3] \\
C V  &$6\ ^3P_2$&$ 1\ ^1S_0$& 32.4086& 32.4103& 1.563[-2]& 1.52[-2]  & 1.304[\hspace{0.35em}3]&  1.27[\hspace{0.35em}3] \\
N VI &$2\ ^3P_2$&$ 1\ ^1S_0$& 29.0807& 29.0815& 7.188[-1]& 7.18[-1]  & 1.031[\hspace{0.35em}5]&  1.03[\hspace{0.35em}5] \\
O VII&$2\ ^3P_2$&$ 1\ ^1S_0$& 21.7998& 21.8044& 5.467[-1]& 5.47[-1]  & 3.311[\hspace{0.35em}5]&  3.31[\hspace{0.35em}5] \\
\multicolumn{9}{c}{E2 transitions} \\
C V  &$3\ ^1D_2$&$ 1\ ^1S_0$& 34.9946& 34.9953& 4.537[-3]& 3.037[-3] & 1.936[\hspace{0.35em}7]&  1.296[\hspace{0.35em}7]\\
C V  &$3\ ^1D_2$&$ 2\ ^1S_0$& 248.447& 248.444& 3.907[\hspace{0.35em}0]& 2.613[\hspace{0.35em}0] & 9.244[\hspace{0.35em}5]&  6.163[\hspace{0.35em}5]\\
C V  &$4\ ^1F_3$&$ 2\ ^1P_1$& 197.034& 197.024& 6.382[-1]& 1.193[\hspace{0.35em}0] & 3.438[\hspace{0.35em}5]&  5.428[\hspace{0.35em}5]\\
N VI &$3\ ^1D_2$&$ 1\ ^1S_0$& 24.9137& 24.9142& 2.449[-3]& 1.649[-3] & 5.715[\hspace{0.35em}7]&  3.847[\hspace{0.35em}7]\\
N VI &$3\ ^1D_2$&$ 2\ ^1S_0$& 174.073& 174.068& 1.913[\hspace{0.35em}0]& 1.285[\hspace{0.35em}0] & 2.681[\hspace{0.35em}6]&  1.801[\hspace{0.35em}6]\\
NVI &$4\ ^1F_3$&$ 2\ ^1P_1$& 136.613& 136.606& 5.366[-1]& 5.708[-1] & 1.804[\hspace{0.35em}6]&  1.919[\hspace{0.35em}6]\\
O VII&$3\ ^1D_2$&$ 1\ ^1S_0$& 18.6365& 18.6377& 1.428[-3]& 9.065[-4] & 1.423[\hspace{0.35em}8]&  9.656[\hspace{0.35em}7]\\
O VII&$3\ ^1D_2$&$ 2\ ^1S_0$& 128.685& 128.685& 1.039[\hspace{0.35em}0]& 7.031[-1] & 6.593[\hspace{0.35em}6]&  4.463[\hspace{0.35em}6]\\
O VII&$4\ ^1F_3$&$ 2\ ^1P_1$& 100.235& 100.237& 2.827[-1]& 3.062[-1] & 4.470[\hspace{0.35em}6]&  4.541[\hspace{0.35em}6]\\
\hline
\end{tabular}
\end{center}
\end{table}

In Table~A, we compare our results for wavelengths ${\lambda}$,
line strengths $S$, and transition rates $A$ for M1, M2, and
E2  transitions for He-like C, N, and O with the
data for these transitions listed in \citet{nist96}.
Our theoretical
wavelengths are in excellent agreement with the
precise experimental wavelengths given in \cite{nist96}; the
differences range from 0.0007\% to 0.005\%.
An exception is the $2\, ^3P_2$ -- $1\, ^1S_0$ transition in O~VII,
where the difference is 0.02\%.
The best agreement for forbidden transition rates ($\leq 3$\%)
occurs for M2 transitions; differences for M1 transitions range
from 0.03\% to 25\%; and the worst agreement (2\% -- 50\%)
occurs for E2 transitions. In this regard, it should
be noted that NIST data for E2 transitions are based on
theoretical results published more than 20 years ago \cite{E2-80}.

We also find excellent agreement (0.1\% -- 0.2 \%) for $2\, ^3S_1$ -- $1\, ^1S_0$ M1 and
$2\, ^3P_2$ - $1\, ^1S_0$ M2 transition rates with  results obtained
from relativistic many-body theory in \cite{PJS95}.

To check our results for transitions from higher $n$ states, we
made a number of second-order RMBPT calculations for M1, M2, and E2
transitions between $n$ = 3 and $n$ = 1, 2 states in O~IX and Si~XIII.
We found that differences between
the two sets of calculations decreased with increasing $Z$,
as expected.
Differences for E2 transitions with $\Delta S = 0$
were about 1\% - 5\% for O~VII and  0.01\% - 1\% for Si~XIII.
For E2 intercombination transitions $\Delta S \neq 0$,
differences increased by factor 10,
primarily because of strong cancellations leading to inaccuracies
in the RMBPT calculations.
For the $3\, ^3S_1$ - $2\, ^3S_1$ and $3\, ^3S_1$ - $2\, ^3S_1$ M1 transitions
we found much larger differences with RMBPT owing to
negative energy state (NES) contributions, which are omitted in the
present calculations but are included in the comparison RMBPT calculations.
Those effects are expected to be large for nonrelativistically forbidden
($n_f \neq n_i$) M1 transitions that conserve total spin according to
the analysis given in Ref.~\cite{M1-98}. We therefore expect poor accuracy
($\approx 10$\%) for the rates of all such transitions.
For other M1 transitions, which change spin or do not change $n$,
the accuracy is much better (1\% -- 2\%).

\section*{RESULTS}
\addcontentsline{toc}{section} {RESULTS}

In Tables~I - VI, we present our
wavelengths ${\lambda}$ and transition
rates $A$.  Empty spaces in the tables signify that no
transition is possible.  Level designations are given in four columns:
the first two columns on the left are for E1 and M2
transitions (even-odd or odd-even) and the next two columns with
level designations are for E2 and M1 transitions (without change
of parity). Tables are organized in order of increasing $n$ of the
upper level, with singlet states first followed by triplet states; they are
further ordered by orbital quantum number $L$. All transitions from a
fixed upper level into lower levels are included except levels
where the transition energy is smaller than 100 cm$^{-1}$.

In Table~VII, we list the sum of the transition
rates $A$ taken from Tables~I - VI. Data are given separately
for each type of transitions (E1, E2, M1, M2) together 
with the sum of all transitions.
The last column of Table~VII lists lifetimes (ps) and the
second column presents energies  (cm$^{-1}$) relative to the
ground state. Results for the 57 possible $n\,^{1,3} L_J$  levels with
$n$ = 2 to 6, $L$ = 0 to 4, and $J$ = 0 to 3 are included
for the  C~V, N~VI, O~VII, Ne~IX, Si~XIII, and Ar~XVII.
As can be seen, lifetimes in heliumlike ions
are almost always dominated by E1 transitions.
It should be noted that only single-photon contributions to
decay rates are included. Therefore, 
rates reported in Table~VII for  $2\, ^3S_0$ states,
which dominantly decay by emission of two photons,
are understimated by many orders of magnitude and should be replaced
by accurate two-photon rates given, for example, in Ref.~\cite{twop:97}.

\section*{CONCLUSION}
\addcontentsline{toc}{section} {CONCLUSION}

We have presented a systematic relativistic CI study of
wavelengths and transition rates  between states with $n\leq6$ in
He-like ions with $Z$ = 6, 7, 8, 10, 14, and 18.
All possible  E1, E2, M1, and M2 transitions  are considered.
Length and velocity form matrix elements for E1 and E2 matrix elements
are in close agreement; the very small differences between these forms
can be explained by omitted negative-energy state contributions.
The accuracy of our energy values is about 0.007\% - 0.025\%. 

\section*{Acknowledgments}
The work of W.R.J. and I.M.S. was supported in part by National
Science Foundation Grant No.\ PHY-01-39928.
U.I.S. acknowledges partial support by Grant No.\ B503968 from 
Lawrence Livermore National
Laboratory.


\begin{thebibliography}{19}
\expandafter\ifx\csname natexlab\endcsname\relax\def\natexlab#1{#1}\fi
\expandafter\ifx\csname bibnamefont\endcsname\relax
  \def\bibnamefont#1{#1}\fi
\expandafter\ifx\csname bibfnamefont\endcsname\relax
  \def\bibfnamefont#1{#1}\fi
\expandafter\ifx\csname citenamefont\endcsname\relax
  \def\citenamefont#1{#1}\fi
\expandafter\ifx\csname url\endcsname\relax
  \def\url#1{\texttt{#1}}\fi
\expandafter\ifx\csname urlprefix\endcsname\relax\def\urlprefix{URL }\fi
\providecommand{\bibinfo}[2]{#2}
\providecommand{\eprint}[2][]{\url{#2}}



\bibitem[{\citenamefont{Johnson et~al.}(2002)\citenamefont{Johnson, Savukov,
  Safronova, and Dalgarno}}]{pap1}
\bibinfo{author}{\bibfnamefont{W.~R.} \bibnamefont{Johnson}},
  \bibinfo{author}{\bibfnamefont{I.}~\bibnamefont{Savukov}},
  \bibinfo{author}{\bibfnamefont{U.~I.} \bibnamefont{Safronova}},
  \bibnamefont{and} \bibinfo{author}{\bibfnamefont{A.}~\bibnamefont{Dalgarno}},
  \bibinfo{journal}{Ap.\ J.\ Supl.} \textbf{\bibinfo{volume}{141}}
  (\bibinfo{year}{2002}), \bibinfo{note}{to be published Aug.\ 2002}.

\bibitem[{\citenamefont{Wiese et~al.}(1996)\citenamefont{Wiese, Fuhr, and
  Deters}}]{nist96}
\bibinfo{author}{\bibfnamefont{W.~L.} \bibnamefont{Wiese}},
  \bibinfo{author}{\bibfnamefont{J.~R.} \bibnamefont{Fuhr}}, \bibnamefont{and}
  \bibinfo{author}{\bibfnamefont{T.~M.} \bibnamefont{Deters}},
  \bibinfo{journal}{J.\ Phys.\ Chem.\ Ref.\ Data, Monograph}
  \textbf{\bibinfo{volume}{7}}, \bibinfo{pages}{154} (\bibinfo{year}{1996}).

\bibitem[{\citenamefont{Drake}(1969)}]{M1-69}
\bibinfo{author}{\bibfnamefont{G.~W.~F.} \bibnamefont{Drake}},
  \bibinfo{journal}{Ap.\ J.} \textbf{\bibinfo{volume}{158}},
  \bibinfo{pages}{1199} (\bibinfo{year}{1969}).

\bibitem[{\citenamefont{Drake}(1971)}]{M1-71}
\bibinfo{author}{\bibfnamefont{G.~W.~F.} \bibnamefont{Drake}},
  \bibinfo{journal}{Phys.\ Rev.\ A} \textbf{\bibinfo{volume}{3}},
  \bibinfo{pages}{908} (\bibinfo{year}{1971}).

\bibitem[{\citenamefont{Kundu and Mukherjee}(1985)}]{M1-85}
\bibinfo{author}{\bibfnamefont{B.}~\bibnamefont{Kundu}} \bibnamefont{and}
  \bibinfo{author}{\bibfnamefont{P.~K.} \bibnamefont{Mukherjee}},
  \bibinfo{journal}{Ap.\ J.} \textbf{\bibinfo{volume}{298}},
  \bibinfo{pages}{844} (\bibinfo{year}{1985}).

\bibitem[{\citenamefont{Plante et~al.}(1995)\citenamefont{Plante, Johnson, and
  Sapirstein}}]{PJS95}
\bibinfo{author}{\bibfnamefont{D.~R.} \bibnamefont{Plante}},
  \bibinfo{author}{\bibfnamefont{W.~R.} \bibnamefont{Johnson}},
  \bibnamefont{and}
  \bibinfo{author}{\bibfnamefont{J.}~\bibnamefont{Sapirstein}},
  \emph{\bibinfo{title}{Advances in Atomic, Molecular and Optical Physics}},
  vol.~\bibinfo{volume}{35} (\bibinfo{publisher}{Academic Press},
  \bibinfo{year}{1995}).

\bibitem[{\citenamefont{Derevianko et~al.}(1998)\citenamefont{Derevianko,
  Savukov, Johnson, and Plante}}]{M1-98}
\bibinfo{author}{\bibfnamefont{A.}~\bibnamefont{Derevianko}},
  \bibinfo{author}{\bibfnamefont{I.~M.} \bibnamefont{Savukov}},
  \bibinfo{author}{\bibfnamefont{W.~R.} \bibnamefont{Johnson}},
  \bibnamefont{and} \bibinfo{author}{\bibfnamefont{D.~R.}
  \bibnamefont{Plante}}, \bibinfo{journal}{Phys.\ Rev.\ A}
  \textbf{\bibinfo{volume}{58}}, \bibinfo{pages}{4453} (\bibinfo{year}{1998}).

\bibitem[{\citenamefont{Charles et~al.}(1989)\citenamefont{Charles, Indelicato,
  de~Billy, Tazi, Briand, Simionovici, Dietrich, Bosch, and Liesen}}]{M1Xe-89}
\bibinfo{author}{\bibfnamefont{P.}~\bibnamefont{Charles}},
  \bibinfo{author}{\bibfnamefont{P.}~\bibnamefont{Indelicato}},
  \bibinfo{author}{\bibfnamefont{L.}~\bibnamefont{de~Billy}},
  \bibinfo{author}{\bibfnamefont{C.}~\bibnamefont{Tazi}},
  \bibinfo{author}{\bibfnamefont{J.-P.} \bibnamefont{Briand}},
  \bibinfo{author}{\bibfnamefont{A.}~\bibnamefont{Simionovici}},
  \bibinfo{author}{\bibfnamefont{D.~D.} \bibnamefont{Dietrich}},
  \bibinfo{author}{\bibfnamefont{F.}~\bibnamefont{Bosch}}, \bibnamefont{and}
  \bibinfo{author}{\bibfnamefont{D.~D.} \bibnamefont{Liesen}},
  \bibinfo{journal}{Phys.\ Rev.\ A} \textbf{\bibinfo{volume}{39}},
  \bibinfo{pages}{3725} (\bibinfo{year}{1989}).

\bibitem[{\citenamefont{Dunford et~al.}(1990)\citenamefont{Dunford, Church,
  Liu, Berry, Raphaelian, Hass, and Curtis}}]{M1Br-90}
\bibinfo{author}{\bibfnamefont{R.~W.} \bibnamefont{Dunford}},
  \bibinfo{author}{\bibfnamefont{D.~A.} \bibnamefont{Church}},
  \bibinfo{author}{\bibfnamefont{C.~J.} \bibnamefont{Liu}},
  \bibinfo{author}{\bibfnamefont{H.~G.} \bibnamefont{Berry}},
  \bibinfo{author}{\bibfnamefont{M.~L.~A.} \bibnamefont{Raphaelian}},
  \bibinfo{author}{\bibfnamefont{M.}~\bibnamefont{Hass}}, \bibnamefont{and}
  \bibinfo{author}{\bibfnamefont{L.~J.} \bibnamefont{Curtis}},
  \bibinfo{journal}{Phys.\ Rev.\ A} \textbf{\bibinfo{volume}{41}},
  \bibinfo{pages}{4109} (\bibinfo{year}{1990}).

\bibitem[{\citenamefont{Simionovici et~al.}(1994)\citenamefont{Simionovici,
  Birkett, Marrus, Charles, Indelicato, Dietrich, and Finlayson}}]{M1Nb-94}
\bibinfo{author}{\bibfnamefont{A.}~\bibnamefont{Simionovici}},
  \bibinfo{author}{\bibfnamefont{B.~B.} \bibnamefont{Birkett}},
  \bibinfo{author}{\bibfnamefont{R.}~\bibnamefont{Marrus}},
  \bibinfo{author}{\bibfnamefont{P.}~\bibnamefont{Charles}},
  \bibinfo{author}{\bibfnamefont{P.}~\bibnamefont{Indelicato}},
  \bibinfo{author}{\bibfnamefont{D.~D.} \bibnamefont{Dietrich}},
  \bibnamefont{and}
  \bibinfo{author}{\bibfnamefont{K.}~\bibnamefont{Finlayson}},
  \bibinfo{journal}{Phys.\ Rev.\ A} \textbf{\bibinfo{volume}{49}},
  \bibinfo{pages}{3553} (\bibinfo{year}{1994}).

\bibitem[{\citenamefont{Schmidt et~al.}(1994)\citenamefont{Schmidt, Forck,
  Grieser, Habs, Kenntner, Miersch, Repnow, Schramm, Sch\"{u}ssler, Schwalm
  et~al.}}]{M1C-94}
\bibinfo{author}{\bibfnamefont{H.~T.} \bibnamefont{Schmidt}},
  \bibinfo{author}{\bibfnamefont{P.}~\bibnamefont{Forck}},
  \bibinfo{author}{\bibfnamefont{M.}~\bibnamefont{Grieser}},
  \bibinfo{author}{\bibfnamefont{D.}~\bibnamefont{Habs}},
  \bibinfo{author}{\bibfnamefont{J.}~\bibnamefont{Kenntner}},
  \bibinfo{author}{\bibfnamefont{G.}~\bibnamefont{Miersch}},
  \bibinfo{author}{\bibfnamefont{R.}~\bibnamefont{Repnow}},
  \bibinfo{author}{\bibfnamefont{U.}~\bibnamefont{Schramm}},
  \bibinfo{author}{\bibfnamefont{T.}~\bibnamefont{Sch\"{u}ssler}},
  \bibinfo{author}{\bibfnamefont{D.}~\bibnamefont{Schwalm}},
  \bibnamefont{et~al.}, \bibinfo{journal}{Phys.\ Rev.\ Lett.}
  \textbf{\bibinfo{volume}{72}}, \bibinfo{pages}{1616} (\bibinfo{year}{1994}).

\bibitem[{\citenamefont{Kundu and Mukherjee}(1989)}]{M2-89}
\bibinfo{author}{\bibfnamefont{B.}~\bibnamefont{Kundu}} \bibnamefont{and}
  \bibinfo{author}{\bibfnamefont{P.~K.} \bibnamefont{Mukherjee}},
  \bibinfo{journal}{Phys.\ Scr.} \textbf{\bibinfo{volume}{39}},
  \bibinfo{pages}{722} (\bibinfo{year}{1989}).

\bibitem[{\citenamefont{Simionovici et~al.}(1993)\citenamefont{Simionovici,
  Birkett, Briand, Charles, Dietrich, Finlayson, Indelicato, Liesen, and
  Marrus}}]{M2Ag-93}
\bibinfo{author}{\bibfnamefont{A.}~\bibnamefont{Simionovici}},
  \bibinfo{author}{\bibfnamefont{B.~B.} \bibnamefont{Birkett}},
  \bibinfo{author}{\bibfnamefont{J.-P.} \bibnamefont{Briand}},
  \bibinfo{author}{\bibfnamefont{P.}~\bibnamefont{Charles}},
  \bibinfo{author}{\bibfnamefont{D.~D.} \bibnamefont{Dietrich}},
  \bibinfo{author}{\bibfnamefont{K.}~\bibnamefont{Finlayson}},
  \bibinfo{author}{\bibfnamefont{P.}~\bibnamefont{Indelicato}},
  \bibinfo{author}{\bibfnamefont{D.}~\bibnamefont{Liesen}}, \bibnamefont{and}
  \bibinfo{author}{\bibfnamefont{R.}~\bibnamefont{Marrus}},
  \bibinfo{journal}{Phys.\ Rev.\ A} \textbf{\bibinfo{volume}{48}},
  \bibinfo{pages}{1695} (\bibinfo{year}{1993}).

\bibitem[{\citenamefont{Godefroid and Verhaegen}(1980)}]{E2-80}
\bibinfo{author}{\bibfnamefont{M.}~\bibnamefont{Godefroid}} \bibnamefont{and}
  \bibinfo{author}{\bibfnamefont{G.}~\bibnamefont{Verhaegen}},
  \bibinfo{journal}{J.\ Phys. B} \textbf{\bibinfo{volume}{13}},
  \bibinfo{pages}{3081} (\bibinfo{year}{1980}).

\bibitem[{\citenamefont{Chen et~al.}(1993)\citenamefont{Chen, Cheng, and
  Johnson}}]{chen93}
\bibinfo{author}{\bibfnamefont{M.~H.} \bibnamefont{Chen}},
  \bibinfo{author}{\bibfnamefont{K.}~\bibnamefont{Cheng}}, \bibnamefont{and}
  \bibinfo{author}{\bibfnamefont{W.~R.} \bibnamefont{Johnson}},
  \bibinfo{journal}{Phys.\ Rev.\ A} \textbf{\bibinfo{volume}{47}},
  \bibinfo{pages}{3692} (\bibinfo{year}{1993}).

\bibitem[{\citenamefont{Cheng et~al.}(1994)\citenamefont{Cheng, Chen, Johnson,
  and Sapirstein}}]{cheng94}
\bibinfo{author}{\bibfnamefont{K.~T.} \bibnamefont{Cheng}},
  \bibinfo{author}{\bibfnamefont{M.~H.} \bibnamefont{Chen}},
  \bibinfo{author}{\bibfnamefont{W.~R.} \bibnamefont{Johnson}},
  \bibnamefont{and}
  \bibinfo{author}{\bibfnamefont{J.}~\bibnamefont{Sapirstein}},
  \bibinfo{journal}{Phys.\ Rev.\ A} \textbf{\bibinfo{volume}{50}},
  \bibinfo{pages}{247} (\bibinfo{year}{1994}).

\bibitem[{\citenamefont{Plante et~al.}(1994)\citenamefont{Plante, Johnson, and
  Sapirstein}}]{PJS94}
\bibinfo{author}{\bibfnamefont{D.~R.} \bibnamefont{Plante}},
  \bibinfo{author}{\bibfnamefont{W.~R.} \bibnamefont{Johnson}},
  \bibnamefont{and}
  \bibinfo{author}{\bibfnamefont{J.}~\bibnamefont{Sapirstein}},
  \bibinfo{journal}{Phys.\ Rev.\ A} \textbf{\bibinfo{volume}{49}},
  \bibinfo{pages}{3519} (\bibinfo{year}{1994}).

\bibitem[{\citenamefont{Johnson et~al.}(1988)\citenamefont{Johnson, Blundell,
  and Sapirstein}}]{JBS88}
\bibinfo{author}{\bibfnamefont{W.~R.} \bibnamefont{Johnson}},
  \bibinfo{author}{\bibfnamefont{S.~A.} \bibnamefont{Blundell}},
  \bibnamefont{and}
  \bibinfo{author}{\bibfnamefont{J.}~\bibnamefont{Sapirstein}},
  \bibinfo{journal}{Phys.\ Rev.\ A} \textbf{\bibinfo{volume}{37}},
  \bibinfo{pages}{2764} (\bibinfo{year}{1988}).

\bibitem[{\citenamefont{Derevianko and Johnson}(1997)}]{twop:97}
\bibinfo{author}{\bibfnamefont{A.}~\bibnamefont{Derevianko}} \bibnamefont{and}
  \bibinfo{author}{\bibfnamefont{W.~R.} \bibnamefont{Johnson}},
  \bibinfo{journal}{Phys.\ Rev.\ A} \textbf{\bibinfo{volume}{56}},
  \bibinfo{pages}{1288} (\bibinfo{year}{1997}).

\end{thebibliography}

\newpage

\section*{EXPLANATION OF TABLES}
\addcontentsline{toc}{section}{EXPLANATION OF TABLES}

\noindent TABLE~I:

\noindent  Wavelengths $\lambda$ (\AA) and radiative
transition rates $A$ (s$^{-1}$) for transitions from $n\,
^{1,3}L_J$ upper level to all $n'\,^{1,3}L^{'}_{J'}$ lower
levels in He-like carbon.
 \vspace{1pc}

\noindent TABLE~II:

\noindent  Wavelengths $\lambda$ (\AA) and radiative
transition rates $A$ (s$^{-1}$) for transitions from $n\,
^{1,3}L_J$ upper level to all $n'\,^{1,3}L^{'}_{J'}$ lower
levels in He-like nitogen.
 \vspace{1pc}

 \noindent TABLE~III:

\noindent  Wavelengths $\lambda$ (\AA) and radiative
transition rates $A$ (s$^{-1}$) for transitions from $n\,
^{1,3}L_J$ upper level to all $n'\,^{1,3}L^{'}_{J'}$ lower
levels in He-like oxygen.
 \vspace{1pc}

 \noindent TABLE~IV:

\noindent  Wavelengths $\lambda$ (\AA) and radiative
transition rates $A$ (s$^{-1}$) for transitions from $n\,
^{1,3}L_J$ upper level to all $n'\,^{1,3}L^{'}_{J'}$ lower
levels in He-like neon.
 \vspace{1pc}

 \noindent TABLE~V:

\noindent  Wavelengths $\lambda$ (\AA) and radiative
transition rates $A$ (s$^{-1}$) for transitions from $n\,
^{1,3}L_J$ upper level to all $n'\,^{1,3}L^{'}_{J'}$ lower
levels in He-like silicon.
 \vspace{1pc}

 \noindent TABLE~VI:

\noindent  Wavelengths $\lambda$ (\AA) and radiative
transition rates $A$ (s$^{-1}$) for transitions from $n\,
^{1,3}L_J$ upper level to all $n'\,^{1,3}L^{'}_{J'}$ lower
levels in He-like argon.
 \vspace{1pc}

\noindent The following notation is used: 
$n\,^{1,3}L_J \equiv (1snl)\ ^{1,3}L_J$ and
we use a[b]=a$10^{b}$ to represent powers of ten. 
\vspace{1pc}

\noindent In the first two columns and the sixth  and seventh
columns, we give the state designations $n\,^{1,3}L_J$ and
$n'\, ^{1,3}L^{'}_{J'}$ for upper and lower levels, respectively.
\vspace{1pc}

\noindent The rows contain wavelengths in \AA ~ and transition
rates in s$^{-1}$ for E1 and M2 transitions followed by wavelengths in
\AA ~ and transition rates in s$^{-1}$ for E2 and M1 transitions.
Empty spaces signify that the corresponding transitions do not occur.
\vspace{1pc}

\noindent TABLE~VII:

\noindent  Energies $E$ (cm$^{-1}$), total radiative E1, E2, M1,
and M2 transition rates $A$ (s$^{-1}$) and lifetimes $\tau$
(10$^{-12}$ s) for  $n\,^{1,3}L_J$ levels in C~V, N~VI, O~VII, 
Ne~IX, Si~XIII, and Ar~XVII. \vspace{1pc}

\noindent In the first column we give the state designations 
$n\,^{1,3}L_J$.
\vspace{1pc}

\noindent The following notation is used: 
$n\,^{1,3}L_J \equiv (1snl)\ ^{1,3}L_J$ and
we use a[b]=a$10^{b}$ to represent powers of ten. 
\vspace{1pc}

\noindent The rows contain energies in cm$^{-1}$, 
radiative E1, E2, M1, M2 rates, total (E1+E2+M1+M2) rates  $A$
(s$^{-1}$), and lifetimes $\tau$ in 10$^{-12}$.

\newpage

\renewcommand{\thetable}{\Roman{table}}
\setcounter{table}{0}

\begin{center}

\topcaption{Wavelengths $\lambda$ (\AA) and
radiative transition rates $A$ (s$^{-1}$) for transitions from $n\ ^{1,3}L_J$
upper level to all
$n'\ ^{1,3}L^{'}_{J'}$ lower levels in He-like carbon.
Numbers in brackets represent powers of 10.\label{tab-c}}
\par
\begin{supertabular}{llllllllll}
\hline
\multicolumn{1}{c}{Upper} &
\multicolumn{1}{c}{Lower} &
\multicolumn{1}{c}{$\lambda$ (\AA)} &
\multicolumn{1}{c}{$A^{\rm E1}$ (s$^{-1}$)} &
\multicolumn{1}{c}{$A^{\rm M2}$ (s$^{-1}$)} &
\multicolumn{1}{c}{Upper} &
\multicolumn{1}{c}{Lower} &
\multicolumn{1}{c}{$\lambda$ (\AA)} &
\multicolumn{1}{c}{$A^{\rm E2}$ (s$^{-1}$)} &
\multicolumn{1}{c}{$A^{\rm M1}$ (s$^{-1}$)\rule{0ex}{2.8ex}} \\
\hline
          &           &        &            &        &$2\ ^1S_0$&$ 2\ ^3S_1$& 2283.0 &            &1.870[-03]\\
[0.3pc]
$2\ ^1P_1$&$ 1\ ^1S_0$& 40.267 &8.862[\hspace{0.35em}11]&          &$2\ ^1P_1$&$ 2\ ^3P_1$& 3540.9 &6.974[-05]  &1.142[-02]\\
$2\ ^1P_1$&$ 2\ ^3S_1$& 1385.5 &7.611[\hspace{0.35em}03]&1.052[-02]&$2\ ^1P_1$&$ 2\ ^3P_0$& 3542.5 &            &1.286[-02]\\
$2\ ^1P_1$&$ 2\ ^1S_0$& 3524.7 &1.673[\hspace{0.35em}07]&          &$2\ ^1P_1$&$ 2\ ^3P_2$& 3558.0 &2.310[-05]  &1.037[-02]\\
[0.3pc]
          &           &        &            &        &$2\ ^3S_1$&$ 1\ ^1S_0$& 41.472 &            &4.874[\hspace{0.35em}01]\\
[0.3pc]
$2\ ^3P_0$&$ 2\ ^3S_1$& 2275.6 &5.670[\hspace{0.35em}07]&          &          &           &        &            &          \\
[0.3pc]
$2\ ^3P_1$&$ 1\ ^1S_0$& 40.730 &2.701[\hspace{0.35em}07]&          &          &           &        &            &          \\
\hline\hline
\end{supertabular}
\end{center}

\newpage

\begin{center}

\topcaption{Wavelengths $\lambda$ (\AA) and
radiative transition rates $A$ (s$^{-1}$) for transitions from $n\ ^{1,3}L_J$
upper level to all
$n'\ ^{1,3}L^{'}_{J'}$ lower levels in He-like nitrogen.
Numbers in brackets represent powers of 10.\label{tab-n}}
\par
\begin{supertabular}{llllllllll}
\hline
\multicolumn{1}{c}{Upper} &
\multicolumn{1}{c}{Lower} &
\multicolumn{1}{c}{$\lambda$ (\AA)} &
\multicolumn{1}{c}{$A^{\rm E1}$ (s$^{-1}$)} &
\multicolumn{1}{c}{$A^{\rm M2}$ (s$^{-1}$)} &
\multicolumn{1}{c}{Upper} &
\multicolumn{1}{c}{Lower} &
\multicolumn{1}{c}{$\lambda$ (\AA)} &
\multicolumn{1}{c}{$A^{\rm E2}$ (s$^{-1}$)} &
\multicolumn{1}{c}{$A^{\rm M1}$ (s$^{-1}$)\rule{0ex}{2.8ex}} \\
\hline
$2\ ^1S_0$&$ 2\ ^3P_1$& 102812&9.193[-02]&          &$2\ ^1S_0$&$ 2\ ^3S_1$& 1870.2&          &6.163[-03]\\
$2\ ^1S_0$&$ 2\ ^3P_2$& 146464&          &1.574[-12]&          &           &       &          &          \\
[0.4pc]
$2\ ^1P_1$&$ 1\ ^1S_0$& 28.786&1.806[\hspace{0.35em}12]&          &$2\ ^1P_1$&$ 2\ ^3P_0$& 2813.4&          &6.249[-02]\\
$2\ ^1P_1$&$ 2\ ^3S_1$& 1135.9&2.391[\hspace{0.35em}04]&1.989[-02]&$2\ ^1P_1$&$ 2\ ^3P_1$& 2814.0&2.647[-04]&5.417[-02]\\
$2\ ^1P_1$&$ 2\ ^1S_0$& 2893.2&2.099[\hspace{0.35em}07]&          &$2\ ^1P_1$&$ 2\ ^3P_2$& 2837.2&8.619[-05]&5.318[-02]\\
[0.4pc]
          &           &       &          &          &$2\ ^3S_1$&$ 1\ ^1S_0$& 29.534&          &2.543[\hspace{0.35em}02]\\
[0.4pc]
$2\ ^3P_0$&$ 2\ ^3S_1$& 1905.2&6.797[\hspace{0.35em}07]&          &          &           &       &          &          \\
\hline\hline
\end{supertabular}
\end{center}

\newpage

\begin{center}

\topcaption{Wavelengths $\lambda$ (\AA) and
radiative transition rates $A$ (s$^{-1}$) for transitions from $n\ ^{1,3}L_J$
upper level to all  
$n'\ ^{1,3}L^{'}_{J'}$ lower levels in He-like oxygen. 
Numbers in brackets represent powers of 10.\label{tab-ox}}
\par
\begin{supertabular}{llllllllll}
\hline
\multicolumn{1}{c}{Upper} &
\multicolumn{1}{c}{Lower} &
\multicolumn{1}{c}{$\lambda$ (\AA)} &
\multicolumn{1}{c}{$A^{\rm E1}$ (s$^{-1}$)} &
\multicolumn{1}{c}{$A^{\rm M2}$ (s$^{-1}$)} &
\multicolumn{1}{c}{Upper} &
\multicolumn{1}{c}{Lower} &
\multicolumn{1}{c}{$\lambda$ (\AA)} &
\multicolumn{1}{c}{$A^{\rm E2}$ (s$^{-1}$)} &
\multicolumn{1}{c}{$A^{\rm M1}$ (s$^{-1}$)\rule{0ex}{2.8ex}} \\
\hline
$2\ ^1S_0$&$2\ ^3P_1$&48318. &1.435[\hspace{0.35em}00]&            &$2\ ^1S_0$&$2\ ^3S_1$&1582.1 &          &1.707[-02]\\
$2\ ^1S_0$&$2\ ^3P_2$&65794. &          &6.360[-11]  &$2\ ^1S_0$&$2\ ^3S_1$&1582.1 &          &1.707[-02]\\
 [0.3pc]
$2\ ^1P_1$&$1\ ^1S_0$&21.600 &3.302[\hspace{0.35em}12]&            &$2\ ^1P_1$&$2\ ^3P_1$&2328.8 &8.106[-04]&2.033[-01]\\
$2\ ^1P_1$&$2\ ^3S_1$&960.83 &6.386[\hspace{0.35em}04]&3.401[-02]  &$2\ ^1P_1$&$2\ ^3P_2$&2359.0 &2.578[-04]&2.100[-01]\\
$2\ ^1P_1$&$2\ ^1S_0$&2446.8 &2.548[\hspace{0.35em}07]&            &$2\ ^1P_1$&$2\ ^3P_0$&2325.7 &          &2.393[-01]\\
[0.3pc]
          &        . &       &          & &           $2\ ^3S_1$&$1\ ^1S_0$&22.097 &  	&1.047[[\hspace{0.35em}03]\\
$2\ ^3P_0$&$2\ ^3S_1$&1637.2 & 7.940[\hspace{0.35em}07]&           &          &          &       &          &         \\
\hline
\end{supertabular}
\end{center}
\newpage

\begin{center}

\topcaption{Wavelengths $\lambda$ (\AA) and radiative transition
rates $A$ (s$^{-1}$) for transitions from $n\ ^{1,3}L_J$ upper
level to all $n'\ ^{1,3}L^{'}_{J'}$ lower levels in He-like
neon. Numbers in brackets represent powers of 10.\label{tab-ne}}
\par
\begin{supertabular}{llllllllll}
\hline \multicolumn{1}{c}{Upper} &
\multicolumn{1}{c}{Lower} &
\multicolumn{1}{c}{$\lambda$ (\AA)} &
\multicolumn{1}{c}{$A^{\rm E1}$ (s$^{-1}$)} &
\multicolumn{1}{c}{$A^{\rm M2}$ (s$^{-1}$)} &
\multicolumn{1}{c}{Upper} &
\multicolumn{1}{c}{Lower} &
\multicolumn{1}{c}{$\lambda$ (\AA)} &
\multicolumn{1}{c}{$A^{\rm E2}$ (s$^{-1}$)} &
\multicolumn{1}{c}{$A^{\rm M1}$ (s$^{-1}$)\rule{0ex}{2.8ex}} \\
\hline
$2\ ^1S_0$&$2\ ^3P_1$&24487. & 2.491[\hspace{0.35em}01] &             &$2\ ^1S_0$&$2\ ^3S_1$&1207.2 &            &  9.149[-02]\\
$2\ ^1S_0$&$2\ ^3P_2$&39521. &            &  4.962[-10] &           &                 &            &            \\
[0.3pc]
$2\ ^1P_1$&$1\ ^1S_0$&13.446 & 8.853[\hspace{0.35em}12] &             &$2\ ^1P_1$&$2\ ^3P_1$&1722.2 & 4.993[-03] &  1.782[\hspace{0.35em}00]\\
$2\ ^1P_1$&$2\ ^3S_1$&730.90 & 3.259[\hspace{0.35em}05] &  8.215[-02] &$2\ ^1P_1$&$2\ ^3P_2$&1769.6 & 1.478[-03] &  1.928[\hspace{0.35em}00]\\
$2\ ^1P_1$&$2\ ^1S_0$&1852.5 & 3.544[\hspace{0.35em}07] &             &$2\ ^1P_1$&$2\ ^3P_0$&1713.4 &            &  2.175[\hspace{0.35em}00]\\
[0.3pc]
          &          &       &            &             & $2\ ^3S_1$&$1\ ^1S_0$&13.698&            &  1.094[\hspace{0.35em}04]\\
[0.3pc]
$2\ ^3P_0$&$2\ ^3S_1$&1274.6 & 1.030[\hspace{0.35em}08] &             &           &                 &            &            \\
\hline
\end{supertabular}
\end{center}

\newpage

\begin{center}

\topcaption{Wavelengths $\lambda$ (\AA) and
radiative transition rates $A$ (s$^{-1}$) for transitions from $n\ ^{1,3}L_J$
upper level to all
$n'\ ^{1,3}L^{'}_{J'}$ lower levels in He-like silicon.
Numbers in brackets represent powers of 10.\label{tab-si}}
\par
\begin{supertabular}{llllllllll}
\hline
\multicolumn{1}{c}{Upper} &
\multicolumn{1}{c}{Lower} &
\multicolumn{1}{c}{$\lambda$ (\AA)} &
\multicolumn{1}{c}{$A^{\rm E1}$ (s$^{-1}$)} &
\multicolumn{1}{c}{$A^{\rm M2}$ (s$^{-1}$)} &
\multicolumn{1}{c}{Upper} &
\multicolumn{1}{c}{Lower} &
\multicolumn{1}{c}{$\lambda$ (\AA)} &
\multicolumn{1}{c}{$A^{\rm E2}$ (s$^{-1}$)} &
\multicolumn{1}{c}{$A^{\rm M1}$ (s$^{-1}$)\rule{0ex}{2.8ex}} \\
\hline
$2\ ^1S_0 $&$2\ ^3P_1$&  14642. &3.952[\hspace{0.35em}02]&           &$2\ ^1S_0$&$ 2\ ^3S_1$& 813.24&          & 1.117[\hspace{0.35em}00]\\
[0.3pc]
$2\ ^1P_1 $&$1\ ^1S_0$&  6.6465 &3.757[\hspace{0.35em}13]&           &$2\ ^1P_1$&$ 2\ ^3P_0$& 1084.0&          & 5.893[\hspace{0.35em}01]\\
$2\ ^1P_1 $&$2\ ^3S_1$&  483.99 &3.776[\hspace{0.35em}06]& 3.247[-01]&$2\ ^1P_1$&$ 2\ ^3P_1$& 1105.2&7.398[-02]& 4.482[\hspace{0.35em}01]\\
$2\ ^1P_1 $&$2\ ^1S_0$&  1195.4 &6.279[\hspace{0.35em}07]&           &$2\ ^1P_1$&$ 2\ ^3P_2$& 1199.9&1.668[-02]& 4.544[\hspace{0.35em}01]\\
[0.3pc]
           &          &         &          &           &$2\ ^3S_1$&$ 1\ ^1S_0$& 6.7391&          & 3.605[\hspace{0.35em}05]\\
[0.3pc]
$2\ ^3P_0 $&$2\ ^3S_1$&  874.41 &1.547[\hspace{0.35em}08]&           &          &           &       &          &           \\
[0.3pc]
$2\ ^3P_1 $&$1\ ^1S_0$&  6.6867 &1.551[\hspace{0.35em}11]&           &$2\ ^3P_1$&$ 2\ ^3P_0$& 56447.&          & 9.901[-02]\\
\hline\hline
\end{supertabular}
\end{center}

\newpage

\begin{center}

\topcaption{Wavelengths $\lambda$ (\AA) and
radiative transition rates $A$ (s$^{-1}$) for transitions from $n\ ^{1,3}L_J$
upper level to all  
$n'\ ^{1,3}L^{'}_{J'}$ lower levels in He-like argon. 
Numbers in brackets represent powers of 10.\label{tab-ar}}
\par
\begin{supertabular}{llllllllll}
\hline
\multicolumn{1}{c}{Upper} &
\multicolumn{1}{c}{Lower} &
\multicolumn{1}{c}{$\lambda$ (\AA)} &
\multicolumn{1}{c}{$A^{\rm E1}$ (s$^{-1}$)} &
\multicolumn{1}{c}{$A^{\rm M2}$ (s$^{-1}$)} &
\multicolumn{1}{c}{Upper} &
\multicolumn{1}{c}{Lower} &
\multicolumn{1}{c}{$\lambda$ (\AA)} &
\multicolumn{1}{c}{$A^{\rm E2}$ (s$^{-1}$)} &
\multicolumn{1}{c}{$A^{\rm M1}$ (s$^{-1}$)\rule{0ex}{2.8ex}} \\
\hline
$2\ ^1S_0$&$ 2\ ^3P_1$&      14462. &  9.627[\hspace{0.35em}02]&              &$2\ ^1S_0$&$ 2\ ^3S_1$&      607.78 &  	  &   7.209[\hspace{0.35em}00]\\
[0.3pc]
$2\ ^1P_1$&$ 1\ ^1S_0$&      3.9478 &  1.071[\hspace{0.35em}14]&              &$2\ ^1P_1$&$ 2\ ^3P_0$&      743.22 &  	  &   7.224[\hspace{0.35em}02] \\
$2\ ^1P_1$&$ 2\ ^3S_1$&      348.46 &  2.364[\hspace{0.35em}07]&  1.052[\hspace{0.35em}00]  &$2\ ^1P_1$&$ 2\ ^3P_1$&      773.04 &  5.969[-01]&   5.033[\hspace{0.35em}02] \\
$2\ ^1P_1$&$ 2\ ^1S_0$&      816.70 &  1.134[\hspace{0.35em}08]&              &$2\ ^1P_1$&$ 2\ ^3P_2$&      932.95 &  8.034[-02]&   3.949[\hspace{0.35em}02] \\
[0.3pc]
          &           &             &            &              &$2\ ^3S_1$&$ 1\ ^1S_0$&      3.9931 &  	  &   4.799[\hspace{0.35em}06] \\
[0.3pc]
$2\ ^3P_0$&$ 2\ ^3S_1$&      656.05 &  2.147[\hspace{0.35em}08]&              &          &           &             &            &              \\
[0.3pc]
$2\ ^3P_1$&$ 1\ ^1S_0$&      3.9681 &  1.782[\hspace{0.35em}12]&              &$2\ ^3P_1$&$ 2\ ^3P_0$&      19262. &  	  &   2.451[\hspace{0.35em}00]  \\
\hline\hline
\end{supertabular}
\end{center}
\newpage
\begin{center}
\topcaption{Energies $E$ (cm$^{-1}$), 
total  radiative E1, E2, M1, and M2 transition rates $A$
(s$^{-1}$) and lifetimes $\tau$ (10$^{-12}$ s) for  $n\,
^{1,3}L_J$ levels in C~V, N~VI, O~VII, Ne~IX, Si~XIII, and 
Ar~XVII. Numbers in brackets represent powers of 10.
\label{tab-life}}
\par
\begin{supertabular}{llllllll}
\hline
\multicolumn{1}{c}{Level} &
\multicolumn{1}{c}{$E$ (cm$^{-1}$)} &
\multicolumn{1}{c}{$A^{\rm E1}$ (s$^{-1}$)} &
\multicolumn{1}{c}{$A^{\rm E2}$ (s$^{-1}$)} &
\multicolumn{1}{c}{$A^{\rm M1}$ (s$^{-1}$)} &
\multicolumn{1}{c}{$A^{\rm M2}$ (s$^{-1}$)} &
\multicolumn{1}{c}{$\Sigma A$ (s$^{-1}$)} &
\multicolumn{1}{c}{$\tau$ (10$^{-12}$ s)\rule{0ex}{2.8ex} }\\
\hline
\multicolumn{8}{c}{He-like carbon}\\
\hline
$2\ ^3S_1$&   2411277 &  0.000[\hspace{0.35em}00] & 0.000[\hspace{0.35em}00] & 4.874[\hspace{0.35em}01] & 0.000[\hspace{0.35em}00] & 4.874[\hspace{0.35em}01] & 2.052[\hspace{0.35em}10]\\
$2\ ^1S_0$&   2455080 &  0.000[\hspace{0.35em}00] & 0.000[\hspace{0.35em}00] & 1.870[-03] & 0.000[\hspace{0.35em}00] & 1.870[-03] & 5.348[\hspace{0.35em}14]\\
$2\ ^3P_1$&   2455210 &  8.366[\hspace{0.35em}07] & 0.000[\hspace{0.35em}00] & 0.000[\hspace{0.35em}00] & 3.962[-04] & 8.366[\hspace{0.35em}07] & 1.195[\hspace{0.35em}04]\\
$2\ ^3P_0$&   2455222 &  5.670[\hspace{0.35em}07] & 0.000[\hspace{0.35em}00] & 0.000[\hspace{0.35em}00] & 0.000[\hspace{0.35em}00] & 5.670[\hspace{0.35em}07] & 1.764[\hspace{0.35em}04]\\
$2\ ^3P_2$&   2455345 &  5.720[\hspace{0.35em}07] & 0.000[\hspace{0.35em}00] & 3.345[-05] & 2.624[\hspace{0.35em}04] & 5.723[\hspace{0.35em}07] & 1.747[\hspace{0.35em}04]\\
$2\ ^1P_1$&   2483451 &  8.862[\hspace{0.35em}11] & 9.285[-05] & 3.465[-02] & 1.052[-02] & 8.862[\hspace{0.35em}11] & 1.128[\hspace{0.35em}00]\\
\hline
\end{supertabular}
\end{center}

\end{document}